# Floquet topological state induced by light-driven band inversion in SnTe

**Authors**: F. Chassot[1,✉], G. Kremer[2], A. Pulkkinen[3], C. Wang[1], J. Krempaský[4], J. Minár[3], G. Springholz[5], M. Puppin[6], J.H. Dil[4,6] and C. Monney[1,✉]

**Affiliations**:

[1]Department of Physics and Fribourg Center for Nanomaterials, Université de Fribourg; Fribourg 1700, Switzerland.

[2]Institut Jean Lamour, UMR 7198, CNRS-Université de Lorraine, Campus ARTEM; Nancy 54011, France.

[3]New Technologies-Research Center, University of West Bohemia; Plzeň 301 00, Czech Republic.

[4]Center for Photon Science, Paul Scherrer Institut; Villigen 5232, Switzerland.

[5]Institut für Halbleiter- und Festkörperphysik, Johannes Kepler Universität; Linz 4040, Austria.

[6]Institute of Physics, Ecole Polytechnique Fédérale de Lausanne; Lausanne 1015, Switzerland.

[✉]e-mail: frederic.chassot@unifr.ch, claude.monney@unifr.ch

**Abstract:** High intensity coherent light can dress matter, realizing new hybrid phases that are not accessible in equilibrium[1,2]. This effect results from the coherent interaction between Bloch states inside the solid and the periodic field of impinging photons which produces hybrid light-matter states called Floquet-Bloch states[3–5] that can alter properties of the solid. Optically inducing a topological state in a semiconductor using so-called Floquet engineering is an exciting prospect[6]. However, it has not been realized, despite its theoretical prediction more than 10 years ago. Here we show that an ultrashort-lived topological state that is absent at equilibrium in the ground state of SnTe can be created with femtosecond light pulses. This occurs when the photoexcitation is similar in energy with the band gap of this polar semiconductor. We observe a concomitant renormalization of the band dispersions that reveals the generation of Floquet states connecting to the topological state. We therefore provide the first direct experimental observation of a Floquet topological state and propose that it is driven by a light-induced band inversion in SnTe. Our discovery opens the way for controlling optically on-demand the topological properties of semiconductors.

**Main text:**

Floquet engineering is a promising route to realize new phases of matter with light, with tailored functionality[7–9]. Time-resolved angle-resolved photoemission spectroscopy (tr-ARPES) directly observes such transient effects with momentum resolution by accessing the out-of-equilibrium electronic structure of the solid in presence of an intense light field. In 2016, this approach has been successfully used to observe Floquet-Bloch states in the semiconductor[10] $Bi_2Se_3$, and to show that the interaction between the thus created bands leads to avoided crossing gaps[11]. Another intense research effort of the last decades has been the search for materials with new topological properties, which promise applications for spintronics and quantum computing. An open question arising from these two fields of research, is whether light could be used to coherently drive a material from a trivial to a topological state, into a phase called Floquet topological insulator (FTI)[12,13]. In 2011, a theoretical scheme to generate such a new state of matter in a direct bandgap semiconductor with a large spin-orbit coupling was proposed, but never realized to date[6].

Here we demonstrate that, by using femtosecond light pulses comparable in energy with the band gap of the semiconductor SnTe, we are able to transiently modify its electronic structure. As shown in Fig. 1, a new topological Dirac cone, absent in the equilibrium state, appears and exists only as long as the light pulse interacts with the crystal. This occurs via the generation of Floquet states across the band gap of SnTe and their hybridization with the bottom of the conduction band (CB). On this time scale the crystal structure is not affected by the photoexcitation, providing conclusive evidence for the creation of the out-of-equilibrium FTI state of matter. Figure 1d illustrates this process schematically. A resonant and coherent photoexcitation creates a Floquet replica of the valence band (VB) which superimposes on the conduction band. This VB Floquet replica hybridizes with the CB, leading to an exchange of band character. The resulting band inversion transforms the equilibrium trivial semiconductor into a non-trivial one.

Notably, the semiconductor SnTe has been investigated during the last decade, because it has been proposed to be a topological crystalline insulator (TCI) protected by mirror symmetries[14,15]. Signatures of linear dispersive states, suggestive of Dirac cones as predicted by theoretical works, were found near the Fermi level[16,17]. However, a delicate point is that SnTe undergoes a structural transition from a high-temperature rocksalt phase to a low-temperature rhombohedral phase, associated with the appearance of a polar, and possibly ferroelectric, state[18,19]. This structural distortion has a direct consequence for the possible existence of a TCI state in SnTe. In the rocksalt phase, a band character inversion between the top of the VB and the bottom of the CB gives rise to the appearance of a Dirac cone at the $\Gamma$ point of the (111) surface[20], also reproduced in our density functional theory (DFT) calculation (Supplementary Fig. 1). In contrast, the rhombohedral structural distortion suppresses the band character inversion, thereby eliminating the topological Dirac cone at $\Gamma$, leaving only trivial surface states, as depicted in the DFT calculation of Fig. 1c. At the same time, the resulting polar crystal structure leads to a lifting of the band degeneracy of the bulk VB and CB, and the surface states according to the Rashba model[21]. We previously showed that this Rashba-type spin splitting, and thus the structural distortion, is persistent up to room temperature, questioning the possible existence of a TCI state in SnTe at any temperature *at equilibrium*[22]. We here demonstrate a route to create a topological state in SnTe even in the structurally distorted phase thanks to Floquet engineering.

***Transient topological state***: We start from the rhombohedrally distorted, and thus topologically trivial state of SnTe. Taking advantage of our previous static ARPES work[22], we easily identify the $\overline{K\Gamma K}$ direction for our time-resolved ARPES experiments at a temperature of 30 K, well below the critical temperature of the transition. The short (100 fs) pump pulse of 0.88 eV photoexcites electrons from the valence band to the region close to the bottom of the conduction band. Subsequently, a 6.3 eV probe pulse is used to photoemit electrons from their original or excited states. Fig. 1a and 1b display time-resolved ARPES spectra taken at $t_0$ (when the pump and probe pulse arrive simultaneously) and at $t_0 + 150$ fs, respectively. A complete measurement as a function of different pump-probe delays is available as a movie in the Supplementary Information.

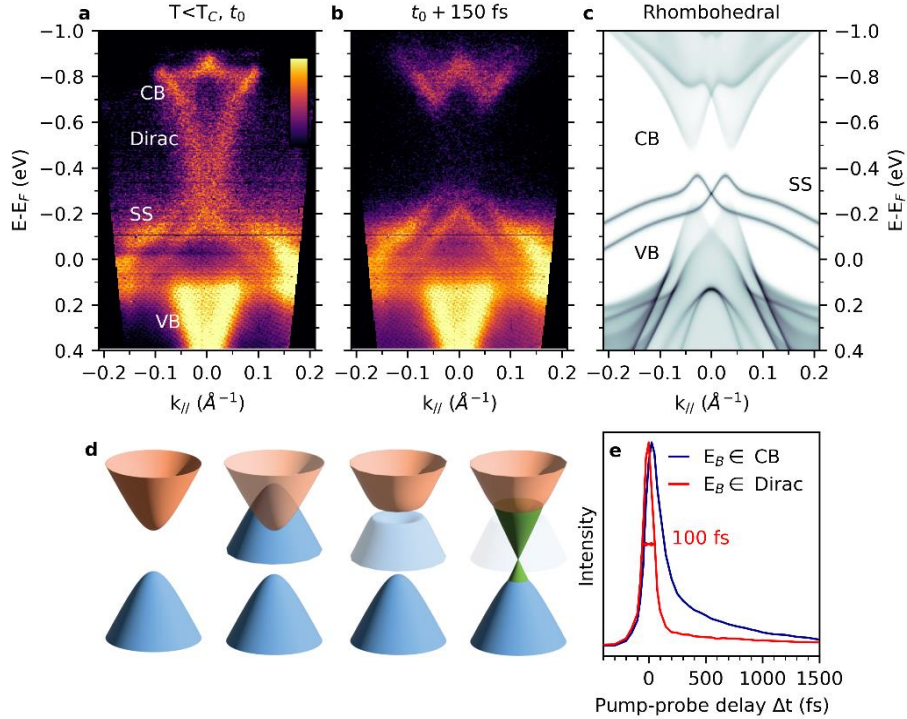

**Fig. 1 | Formation of an ultrashort-lived topological state in polar SnTe through Floquet engineering. a,** Time-resolved ARPES measurement of SnTe(111) along $\overline{K\Gamma K}$ using a probe of 6.3 eV and a pump of 0.88 eV that arrive simultaneously, at time $t_0$. A Dirac cone is observed. **b,** 150 fs after $t_0$, the Dirac cone has disappeared. **c,** SPR-KKR DFT calculations of rhombohedral SnTe(111) using a semi-infinite stack of layers to calculate the bulk and surface states. In this phase, no Dirac cone is expected at equilibrium, in contrast to the experimental data in (**a**). **d,** Schematic illustrations of a Floquet topological insulator, with an initial VB (CB) in blue (orange) that is replicated by the dressing due to the time-periodic electric field of the pump. The replica interacts with the CB (middle) and, through the hybridization, create the necessary condition for the appearance of a Dirac cone. **e,** Intensity as a function of pump-probe delay integrated over the whole momentum range and over 0.1 eV centered at -0.85 eV (in the CB) and at -0.45 (Dirac cone). The Dirac cone exists in presence of the pump pulse only.

Several bands are visible in the data at $t_0$, including a trivial surface state (SS), a bulk VB and the transiently populated CB (the corresponding electronic structure is given on a larger energy scale in Supplementary Fig. 2). However, the most striking feature is that, despite being in the rhombohedral phase, a Dirac cone is crossing the band gap, at a similar position as the topological Dirac cone in the rocksalt structure[23] (see also Supplementary Discussion and Supplementary Fig. 3). At longer timescales, after the photoexcitation pulse is gone, the Dirac cone disappears, and the CB exhibits the 'W' shape predicted by DFT (see Fig. 1c). The time evolution of the transient Dirac cone is shown in Fig. 1e and clearly displays a different behavior compared to the electrons excited to the bottom of the CB. While these electrons remain in the CB for several hundred femtoseconds, the transient Dirac cone has a duration of about 100 fs only. This is very unusual, as electrons in many topological systems typically remain in the Dirac cone for much longer due to a bottleneck effect[24–26]. This sharp and symmetrical intensity profile is therefore as short as it can be, purely limited by our experimental time resolution (see Supplementary Fig. 4 for a measurement of our time-resolution). The electronic structure giving rise to the Dirac cone exists in the presence of the photoexcitation pulse *only*.

*Not a structural transition*: The ultrashort duration of the Dirac cone indicates that its origin is not due to a structural change, which typically occurs on a longer time scale[27–29]. In our experiment, we can actually access the time evolution of the structural distortion through the Rashba splitting of the bulk bands and we do not see any significant change within the first 150 fs after $t_0$, proving the absence of notable atomic displacement (see Supplementary Fig. 5). This further confirms the coexistence of the rhombohedral distortion with a topological state induced by an intense ultrashort light pulse at $t_0$.

*Floquet renormalization:* Concomitantly to the appearance of the Dirac cone at $t_0$, we observe that the CB displays an anomalous flat dispersion. In Fig. 2, a zoom of our ARPES data on the transiently occupied CB is shown near $t_0$ and compared to later times. It can be seen that the photoemission intensity is particularly pronounced along a flat band at 0.8 eV above $E_F$ during the first 100 fs and then, at later times, is distributed more towards lower binding energies, emphasizing the W-shaped CB bottom, expected from DFT (see Fig. 1c). We further characterize the duration of this effect by plotting energy distribution curves (EDCs) across the side of the CB at different time delays, which yields two further hints on what is happening. First, at around 0.8 eV above $E_F$, the EDC can be decomposed in two peaks, rather than being a continuous electron distribution that changes its shape as a function of time, as would be expected for an excited population of electrons in a single band and is observed for an excitation energy of 1.55 eV (Supplementary Fig. 6). Second, the upper peak (near 0.85 eV above $E_F$), which comes from the flat CB, exists only for very short time delay, up to 100 fs, as also visible from the difference plot (Supplementary Fig. 7). Altogether, we observe that the transient flat band appears during photoexcitation only, indicating a renormalization of the CB dispersion due to hybridization with Floquet-Bloch sidebands.

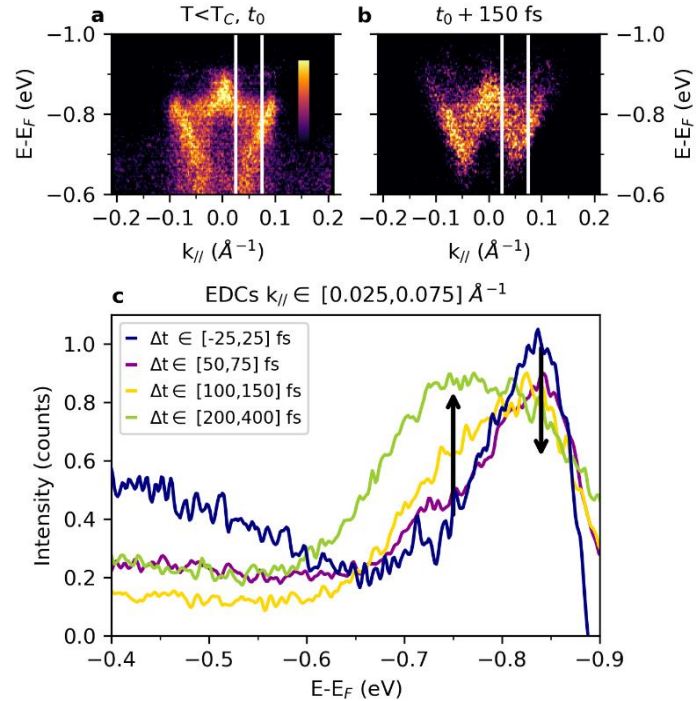

**Fig. 2 | Floquet renormalization of the conduction band**. **a,** Time-resolved ARPES spectra of SnTe(111) using a pump of 0.88 eV, integrated between -25 and 50 fs of pump-probe delays. **b,** Same integrated over 100 to 200 fs. **c,** EDCs for different pump-probe delays integrated between the two white lines in (**a**) and (**b**). The flat band that appears at $t_0$ is due to Floquet states and, later on, it is replaced by the original 'W'-shaped conduction band. The black arrows indicate their evolution in time.

*Fluence and photon energy dependence*: Having shown that a Dirac cone can be photoinduced at $t_0$ by an ultrashort light pulse, we now provide evidence that this effect occurs only close to an optical resonance between the extrema of the bulk VB and CB. In Fig. 3a the normalized intensity of the Dirac cone versus the pump photon energy is shown, revealing several key points. Firstly, different linear polarizations (s and p) yield similar results, indicating that the electric field direction of the probe does not play a major role and thus no laser assisted photoemission effect (LAPE) takes place[11,30]. Secondly, the pronounced maximum of the effect occurs at a photoexcitation energy of 0.88 eV, resonant with the transition near the extrema of the VB to the CB at the $\bar{\Gamma}$ point. For off-resonance excitation with pump energies of 1.03 and 0.95 eV, replicas of the VB shifted by the pump energy are present, but do not show the strong hybridization with the CB leading to a flat band (Supplementary Fig. 8). These points provide further evidence of a Floquet-induced topological state similar to the proposed theoretical mechanism, where a maximal hybridization occurs when the replica and conduction band overlap[6].

The fluence-dependency at a pump photon energy of 0.88 eV, shown in Fig. 3b indicates that the lifetime of the Dirac cone remains unaffected by the fluence, i.e. it is independent of the deposited energy, or the total excited electron population. Furthermore, the inset of Fig. 3b shows that the evolution of the absolute intensity as a function of fluence displays a clear linear behavior. This linearity confirms that the effect scales quadratically with the incoming electric field and converges to zero without photoexcitation. It is thus reminiscent of the optical Stark effect as a consequence of the generation of Floquet states driven by the optical field of the photoexcitation[31].

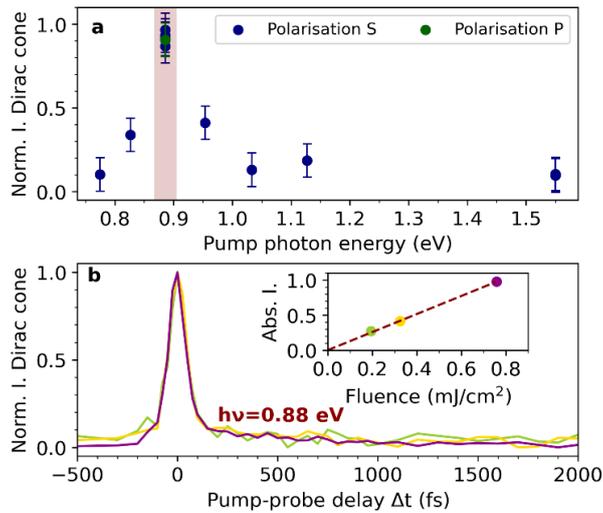

**Fig. 3 | Investigation of the Dirac cone as a function of fluence and pump photon energy. a,** Evolution of the intensity of the Dirac cone as a function of pump photon energy, where the maxima of the signal of the Dirac cone (integrated over E ∈ [−0.35, −0.45] eV and $k_{//} \in$ [−0.03,0.03] Å$^{-1}$) is divided by the maxima of a signal of the surface state to normalize by the amount of absorbed energy. A resonance is observed near the energy of the band gap. **b,** Normalized intensity as a function of pump probe delay for different fluences for a signal integrated over the same region as (A) for a pump energy of 0.88 eV; **b inset,** Absolute intensity of the Dirac cone as a function of fluence for same parameters. No finite lifetime is observed.

*Discussion and outlook*: Using the cartoon depicted in Fig. 4, we propose a scenario based on Floquet physics and exchange of orbital character to explain the mechanism leading to the ultrashort-lived topological state. We recall here that our experiment is performed at low temperature, such that both the VB and the CB are Rashba split. The photoexcitation, similar in energy with the band gap, creates a Floquet replica of the VB on top of the CB bottom (see

Fig. 4a). The interaction between these two bands leads to a subsequent hybridization[32] (see Fig. 4b and Supplementary Fig. 8), which has two consequences. First, it renormalizes the CB dispersion into a flat CB (see Fig. 2). Second, it creates an exchange of orbital character between the top of the VB and the bottom of the CB. This is depicted with the color coding in Fig. 4c. This hybridization introduces a partial Te texture, i.e. a change of the orbital character at the bottom of the CB. We propose that this alters the topological invariant of SnTe, allowing for the transient appearance of the Dirac cone at the Γ point, as long as the Floquet states exist[20].

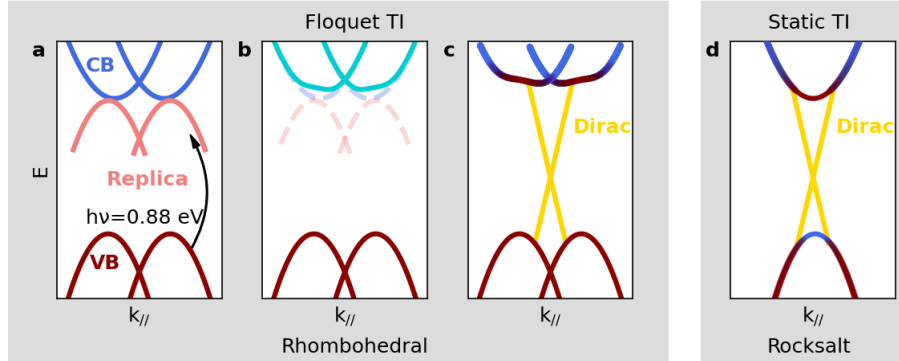

**Fig. 4 | Floquet replica and hybridization of the bands to explain the topological transition. a,** Illustration of the Rashba-split VB (dark red) and CB (dark blue) at equilibrium, in the rhombohedral structure. Upon pumping, a Floquet replica of the VB (light red) is created close to the CB. **b,** The interaction between the replica and the CB leads to a renormalization of the CB (light blue) and (**c**) to an exchange of orbital characters, due to the hybridization, which also changes the topological invariant and create the Dirac cone. **d,** At equilibrium, in the rocksalt structure, a Dirac cone is expected in the presence of a band inversion between the VB and CB[20].

A striking result is the coexistence between maximally Rashba-split bands and a Dirac cone that appears in the presence of the photoexcitation only. The former is due to the rhombohedrally distorted lattice structure, which breaks inversion symmetry in the bulk. This is fundamentally different from the situation at equilibrium, for which a Dirac cone is not allowed in the presence of a large lattice distortion and would typically be present for a rocksalt structure[20] (see Fig. 4d). This is the reason why no Dirac cone is observed after 150 fs, in the absence of light, despite the relatively long occupation of the CB bottom (about 1 ps, see Fig. 1e). This means that we have generated with light a *new out-of-equilibrium state of matter*, for which a topological state exists in the presence of inversion symmetry breaking in SnTe. This is the essence of the FTI state.

These results raise the question of why such physics can be realized in the polar semiconductor SnTe and has not been observed in any other system despite substantial efforts. Especially since, in comparison to other cases of Floquet physics[7,10,32–34], for which high photoexcitation fluences at low photon energies were used, the experimental conditions here, involving relatively mild fluences and high photon energies, might be surprising. The resonance condition of near-bandgap photoexcitation, at the origin of the optical Stark effect, enhances such an effect at low fluence, as evidenced by the linear dependency in fluence (see Fig. 3b)[31]. Moreover, we speculate that the polar nature of SnTe makes it particularly responsive to the dipolar field of light. Finally, although SnTe is not in a TCI phase at low temperature, it is very close to a topological phase transition and therefore highly susceptible to light-induced transformation. Our results thus not only provide the first momentum-resolved experimental verification of a Floquet topological insulator, but also a route to its realization in other systems based on efficient light-matter interaction in proximity to a topological phase transition.

## Methods

### Sample growth

Epitaxial SnTe(111) films of 2 μm thickness were grown by molecular beam epitaxy on BaF$_2$ substrates under ultra-high vacuum (UHV) conditions at a substrate temperature of 350° C and a compound effusion cell. During growth, the SnTe(111) surface exhibits a perfect two-dimensional reflection high-energy electron diffraction pattern revealing a perfect 2D growth mode. After growth, the samples were transferred to the ARPES setup without breaking UHV conditions using a battery-operated vacuum suitcase having a pressure of better than $10^{-10}$ mbar. It is noted, that due to the high density of native Sn vacancies, SnTe intrinsically exhibits a high p-type carrier concentration of typically $2 * 10^{20}$ cm$^{-3}$. For this reason, the Fermi level is always inside the topmost valence band. At room temperature, the lattice parameter of the SnTe layers was determined to be a = 6.325 Å (rhombohedral lattice parameter of 4.472 Å), which is in good agreement with literature values[18]. In our previous paper, we characterized the SnTe epilayer with high resolution x-ray diffraction to prove the perfect (111) orientation of the lattice and the high quality of our samples[22,35].

### (TR-)ARPES measurement

Time and angle-resolved photoemission spectroscopy (TR-ARPES) investigation were carried out using a Scienta DA30 photoelectron analyzer with a base pressure better than $3 \times 10^{-11}$ mbar. The photon sources used for TR-ARPES are based on a femtosecond laser (Pharos, Light Conversion, operating at 1030 nm). Half of its power is converted into 780-nm light with an optical parametric amplifier, which is then frequency-quadrupled to 6.3 eV in β-BaB$_2$O$_4$ crystals to generate probe pulses[36]. The remaining half of the fundamental laser power is directed into a collinear optical parametric amplifier (Orpheus, Light Conversion) to generate pump pulses between 800 nm and 1600 nm (1.55 eV - 0.77 eV) with a duration of about 50 fs. The total temporal resolution was estimated to be equal or better than 100 fs by measuring the width of the cross-correlation between the pump and the probe pulses, using laser-assisted photoemission signal on a reference sample of Bi$_2$Se$_3$. We used s- and p-polarization for the probe and an incident angle of 55° and acquired the data with a negative bias of -5 V. The total energy resolution was about 30 meV and the angular resolution was 0.1°. The samples were cooled to 30 K at rates <5 K/min to avoid thermal stress.

## DFT calculations

The fully relativistic density functional theory calculations of the SnTe(111) surface were performed within the screened Korringa-Kohn-Rostoker (KKR) method in the SPRKKR package[37]. The Te-terminated SnTe(111) surface was modeled as a semi-infinite stack of layers in the atomic sphere approximation (ASA), with basis set truncated at lmax = 2. Exchange-correlation effects were treated at the level of local spin density approximation (LSDA) using the parametrization of Vosko, Wilk and Nusair[38]. The lattice parameters in the hexagonal basis were a=4.528 Å, c=11.090 Å for the rocksalt structure and a=4.523 Å, c=11.227 Å, zTe=0.483 for the rhombohedral structure. The surface-projected electronic band structures, showing kz-integrated bulk bands and the surface states of the SnTe(111) surfaces, are represented by the Bloch spectral function.

## Supplementary Information

### Supplementary Methods

Normalization procedure

In this paper, we employ various normalization procedures tailored to the specific comparisons or highlights we aim to achieve. For the time-resolved band structures (e.g., Fig. 1a-1b), each energy distribution curve (EDC) is divided by its maximum value. This enables us to simultaneously plot both the valence band (VB) and conduction band (CB), despite the differing signal intensities.

In Figures 1e and 3b, the intensity integrated within a defined region is normalized by its maximum value to compare the lifetimes of different states. For the inset of Figure 3B, we plot the maximum absolute intensity of the Dirac cone signal, subtracting the signal before $t_0$ for background removal.

In Figure 3a, to compare the occupation of the topological state under varying fluences and wavelengths, we normalize the maximum intensity of the Dirac cone by the maximum intensity of the surface states above the Fermi level (integrated over $E \in [-0.15, -0.25]$ eV and $k_{//} \in [-0.2, 0.2]$ Å$^{-1}$). This signal serves as the closest approximation to the amount of absorbed energy.

### Supplementary Discussion

TR-ARPES data with 1.55 eV pump photons

Supplementary Fig. 3 shows the TR-ARPES spectra of SnTe(111) measured at 30 K pump with a photon energy of 1.55 eV with an incident fluence of 0.62 mJ/cm$^2$. The data at $t_0$ shows no Dirac cone, indicating that we do not induce a Floquet topological insulator in that case. It is however worth to note that at longer timescale, we observe indication of a Dirac cone in the EDCs. Because of the temporal delay between $t_0$ and the appearance of the effect, we attribute this Dirac cone to a topological transition induced by heat transfer induced by the pump. This agrees with the reduction of the Rashba splitting at the same timescale measured with every pump photon energy, see e.g. Supplementary Fig. 5, and is in agreement with previous literature[23].

# Supplementary figures

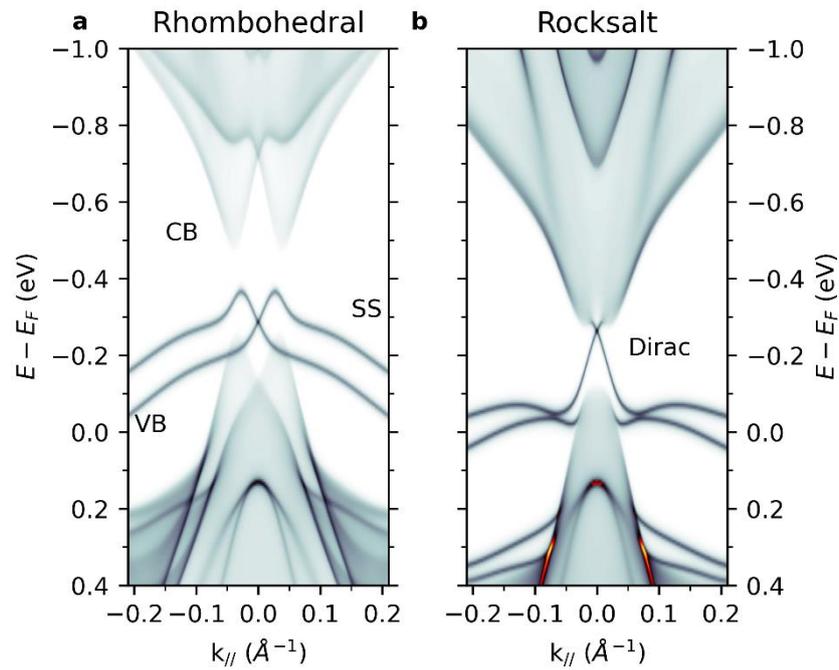

**Supplementary Fig. 1 | SPRKKR DFT** calculation of SnTe along $\overline{K\Gamma K}$ for the rhombohedral (**a**) and rocksalt (**b**) structure to highlight the difference: The structural distortion in the polar rhombohedral structure splits the bulk band in (**a**), whereas the mirror symmetry in the rocksalt structure protects a topological state in (**b**) due to the band inversion.

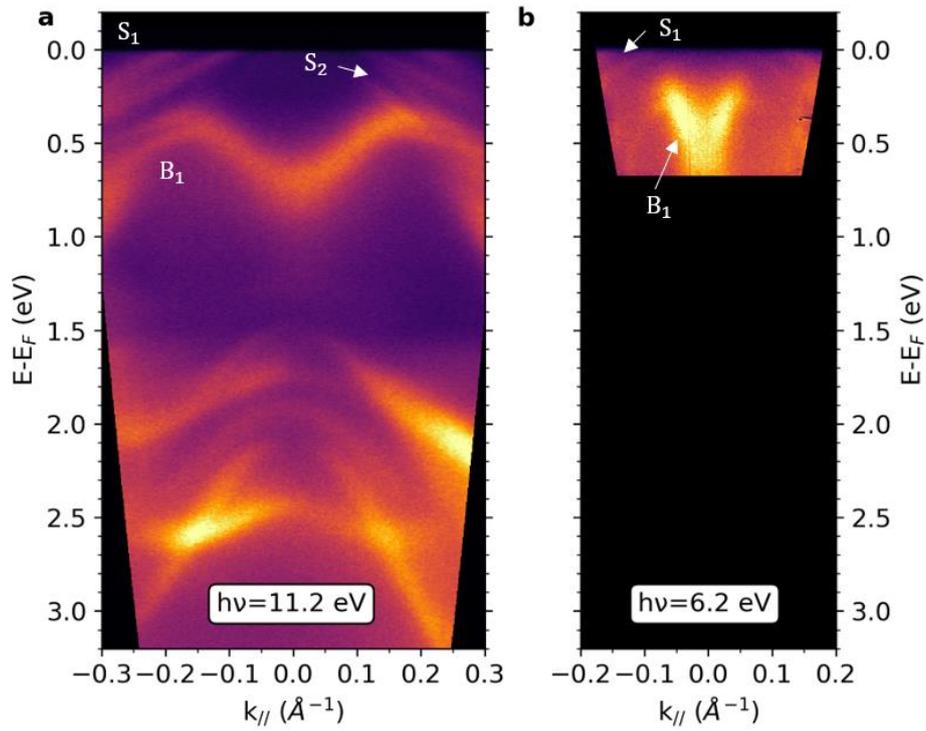

**Supplementary Fig. 2 | Occupied band structure.** ARPES measurements along the $\overline{K\Gamma K}$ high symmetry line at 30 K with a photon energy of (**a**) 11.2 eV and (**b**) 6.2 eV with highlighted surface state ($S_1$), surface resonance state ($S_2$), and bulk state ($B_1$) ; the latter being split because of the break of inversion symmetry in the polar structure.

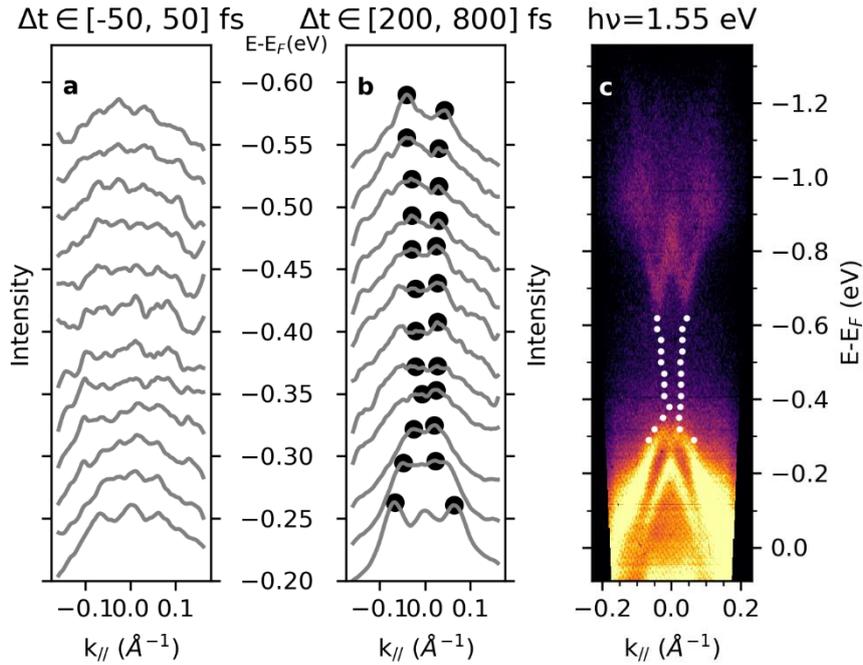

**Supplementary Fig. 3 |** Angular distribution curves at their respective energy integrated over 0.1 eV in time range of $\Delta t \in [-50, 50]$ fs (**a**) and $\Delta t \in [200, 800]$ fs (**b**). **c,** TR-ARPES spectra of SnTe(111) using a probe of 6.3 eV and a pump of 1.55 eV, integrated between 200 and 800 fs of pump-probe delays. This plot clearly shows that the Dirac cone appears only at longer timescale, indicating a structural transition induced by heat transfer from the pump.

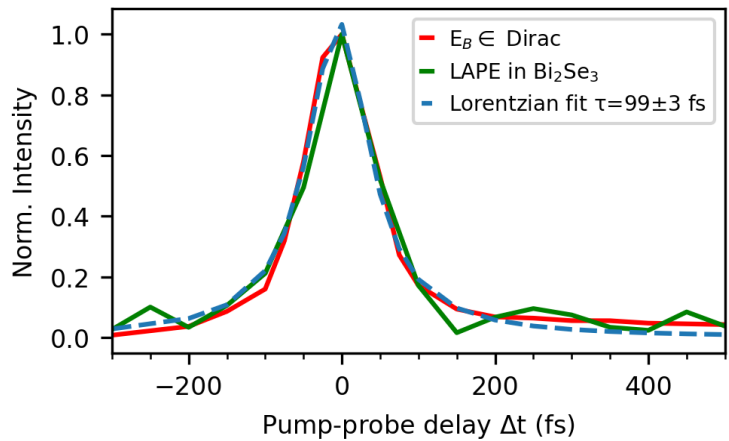

**Supplementary Fig. 4 | Comparison between the time resolution of our experiments and the duration topological state.** Red line: Normalized intensity at the Dirac point (integrated in a box over the whole momentum range centered at -0.45 eV for the data pump with 0.88 eV) plotted as a function of pump-probe delays. Green line: LAPE signal from a reference sample of $Bi_2Se_3$ with the same experimental conditions (the FWHM of the LAPE signal is known in the literature to depend solely on the pump-probe cross correlation) and a Lorentzian fit on the topological signal in blue.

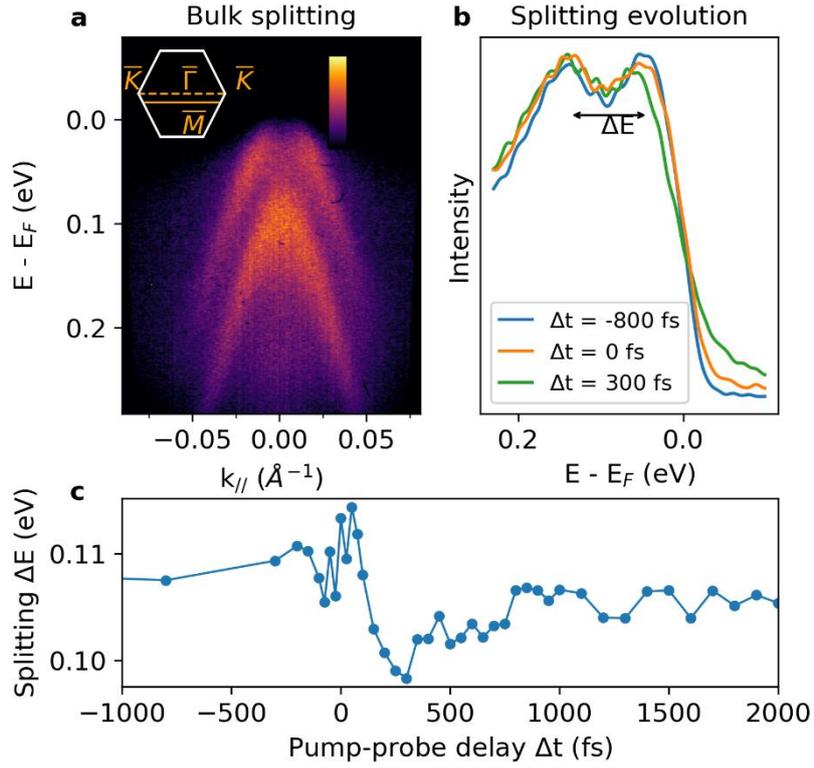

**Supplementary Fig. 5 | Evolution upon pumping of the bulk Rashba splitting as a function of pump-probe delay to prove the absence of a structural transition at $t_0$. a,** ARPES measurement taken along $\overline{K\Gamma K}$ but shifted by $k_{//} = 0.13$ Å$^{-1}$ towards $\overline{M}$ (see continuous line in the Brillouin zone) where the bulk band is split because of the rhombohedral transition ; **b,** EDCs as a function of pump-probe delay measured at $k_{//} \in [-0.02, 0.01]$ Å$^{-1}$ ; **c,** Time evolution of the bulk splitting extracted from a fitting procedure.

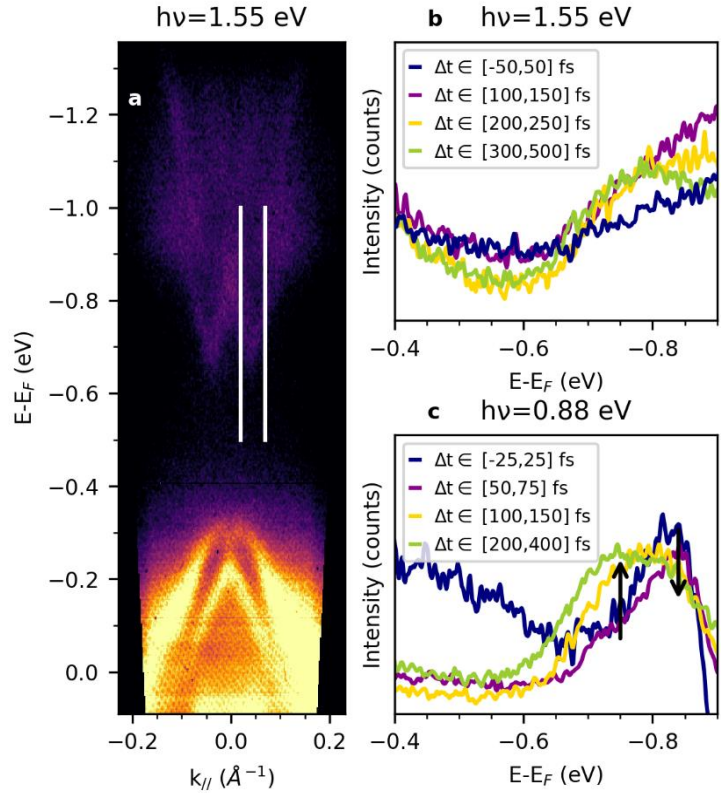

**Supplementary Fig. 6 | Change of the conduction band shape. a,** TR-ARPES spectra of SnTe(111) using a probe of 6.3 eV and a pump of 1.55 eV, integrated between -25 and 800 fs of pump-probe delays ; **b,** EDCs for different pump-probe delays integrated between the two white lines in (**a**) showing no band renormalization but only deexcitation of electrons from the top to the bottom of the CB at 1.55 eV of photon energy ; **c,** same as in (**b**) but for a pump photon energy of 0.88 eV indicating the renormalization of the CB.

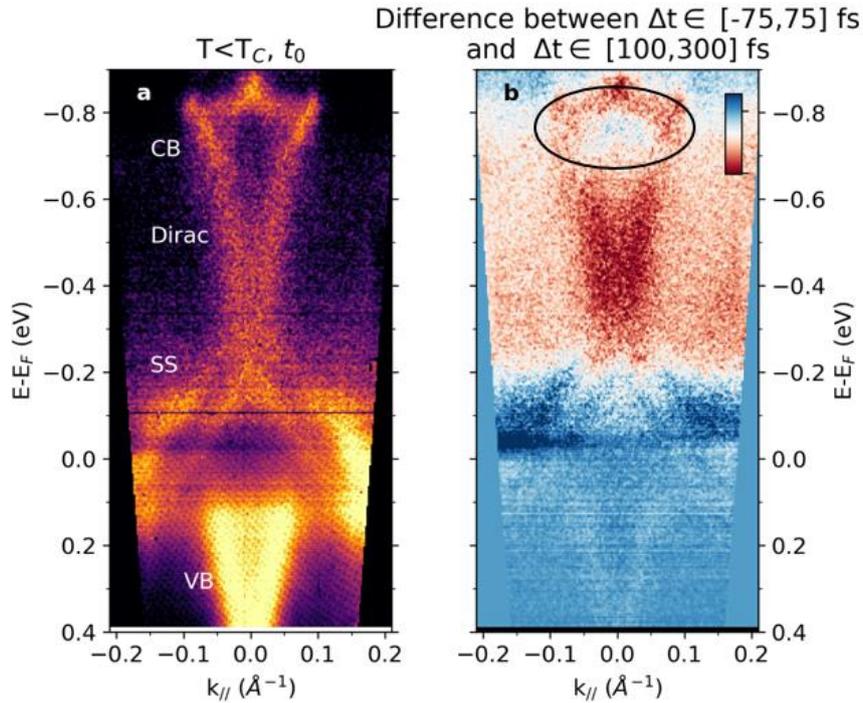

**Supplementary Fig. 7 | Detection of an unexpected and ultrashort topological state in polar SnTe upon pumping and change in the Conduction Band. a,** TR-ARPES measurement of polar SnTe(111) using a probe of 6.3 eV and a pump of 0.88 eV that arrive simultaneously on the sample along $\overline{K\Gamma K}$ ; **b,** Difference between the TR-ARPES measurement integrated between [-75, 75] fs and [100, 300] fs (red : more signal at $t_0$ ; blue more signal after $t_0$). The difference spectrum indicates a change in the conduction band (highlighted in black) going from a flat to a 'W' shape.

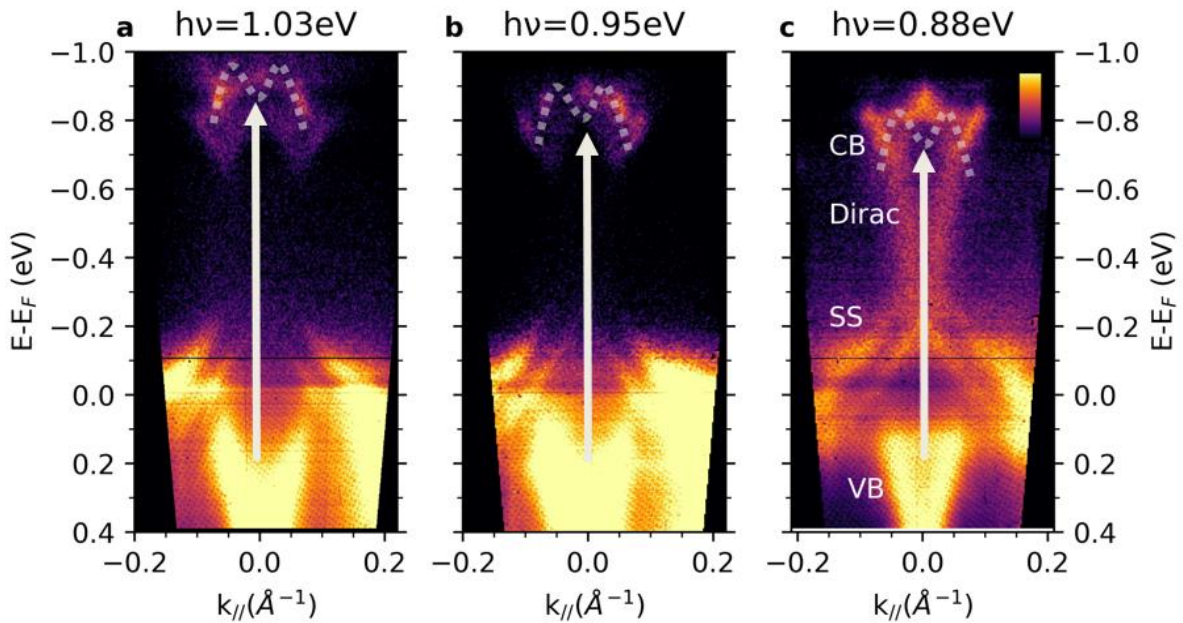

**Supplementary Fig. 8 | Replica**. TR-ARPES spectra of SnTe(111) obtained using a probe of 6.3 eV arriving synchronously with a pump of 1.03 (**a**), 0.95 (**b**) and 0.88 eV (**c**). White arrows indicate the pump photon energy shifting the valence band up, creating replica as indicated by dashed lines that serve as guides to the eye. The replicas are visible in the raw data in (**a**) and (**b**) but hidden in (**c**) underneath the Dirac cone and the hybridized CB.

## Additional information

**Acknowledgements:** We are very grateful to M. Schüler, T. Bzdusek and A. Iliasov for discussions. Technical assistance was provided by F. Bourqui, B. Hediger, S. Falk, M. Baeriswyl and M. Audrey. This project was supported by Swiss National Science Foundation Grant No. 10000782. This work was also supported by the project Quantum materials for applications in sustainable technologies (QM4ST), funded as Project No. CZ.02.01.01/00/22_008/0004572 by Programme Johannes Amos Comenius, call Excellent Research. G.S. acknowledges funding from the Austrian Science Funds, Grant No. 10.55776/PIN6540324 and I-4493, and JKU-Linz Grant 296/LIT-2022-11-SEE-131.

**Author contributions:** F.C., G.K. and C.M. designed the research. F.C. carried out all the experiments and measurements, with assistance of C.W. and C.M.. A.P. and J.M. performed the DFT calculations. G.S. performed sample growth and initial characterization. F.C. and C.M analyzed the data with help from all the authors. F.C. and C.M wrote the paper, with inputs from all authors.